\documentclass[letterpaper, superscriptaddress, twocolumn, showpacs, prb, nofootinbib]{revtex4-1}

\usepackage[]{graphicx}
\usepackage{amsmath}
\usepackage{amsfonts}
\usepackage{color}
\usepackage{hyperref}
\usepackage{bm}
\usepackage{graphicx}
\usepackage{amsbsy}
\usepackage{epstopdf}
\usepackage{animate}
\usepackage{rotating}

\newcommand{\1}{{\openone}}

\definecolor{gold}{rgb}{0.85,.66,0}
\definecolor{brown}{rgb}{0.647,0.165,0.165}

 
\date{\today}

\begin{document}

\title{Optical Cryocooling of Diamond}

\author{M. Kern}
\affiliation{Ulm University, 89081 Ulm, Germany}
\affiliation{Department of Physics and Astronomy, Macquarie University, Sydney, NSW 2109, Australia}
\author{J. Jeske}
\affiliation{Chemical and Quantum Physics, School of Applied Sciences, RMIT University, Melbourne 3001, Australia}
\author{D.M.W. Lau}
\affiliation{Chemical and Quantum Physics, School of Applied Sciences, RMIT University, Melbourne 3001, Australia}
\affiliation{ARC Centre of Excellence for Nanoscale BioPhotonics, RMIT University, Melbourne 3001, Australia}
\author{A.D. Greentree}
\affiliation{Chemical and Quantum Physics, School of Applied Sciences, RMIT University, Melbourne 3001, Australia}
\affiliation{ARC Centre of Excellence for Nanoscale BioPhotonics, RMIT University, Melbourne 3001, Australia}
\author{F. Jelezko}
\affiliation{Institute of Quantum Optics, Ulm University, 89081 Ulm, Germany}
\affiliation{Center for Integrated Quantum Science and Technology (IQST), Ulm University, 89081 Ulm, Germany}
\author{J. Twamley}
\affiliation{Centre for Engineered Quantum Systems, Department of Physics and Astronomy, Macquarie University, Sydney, NSW 2109, Australia}

\begin{abstract}
The cooling of solids by optical means only using anti-Stokes emission has a long history of research and achievements. Such cooling methods have many advantages ranging from no-moving parts or fluids through to operation in vacuum and may have applications to cryosurgery. However achieving large optical cryocooling powers has been difficult to achieve except in certain rare-earth crystals. Through study of the emission and absorption cross sections we find that diamond, containing either NV or SiV (Nitrogen or Silicon vacancy), defects shows potential for optical cryocooling and in particular, NV doping shows promise for  optical refrigeration. We study the optical cooling of doped diamond microcrystals  ranging $10-250\,\mu$m in diameter trapped either in vacuum or in water. For the vacuum case we find  NV-doped microdiamond optical cooling below room temperature could exceed $|\Delta T|>10\, {\rm K}$, for irradiation powers of $P_{in}< 100$ mW. We predict that such  temperature changes should be easily observed via large alterations in the diffusion constant for optically cryocooled microdiamonds trapped in water in an optical tweezer or via spectroscopic signatures such as the ZPL width or Raman line.
\end{abstract}

\maketitle


%

Optical refrigeration, or optical cryocooling, uses anti-Stokes emission in solids to deplete the phonon population within the solid, cooling it. First suggested by Pringsheim \cite{Pringsheim:1929uy}, and initially demonstrated by Epstein {\it et al} in 1995 \cite{Epstein:1995wo}, the phenomena is observed when low entropy laser light is primarily absorbed at wavelengths slightly longer than the  mean fluorescence wavelength ($\lambda_f$) of the material. The light is reemitted with a broadband fluorescence possessing a mean energy which is higher than the incident pump laser.    The increase in energy is due to absorption of lattice phonons (vibrations), reducing the net temperature of the solid. Optical refrigeration has experimentally achieved  temperatures of $T\sim$155K (from room temperature) using ytterbium-doped fluoride crystal ($\rm YLiF_4:Yb^{3+}$) \cite{Seletskiy:2010fk},  $T\sim$250K (from 290K), for the semiconductor CdS \cite{Zhang:2013hy}, and  $T\sim$114K more recently using a multi-pass setup with  Yb-doped crystals \cite{Melgaard:2014eo}. These temperatures outperform Peltier/thermoelectric coolers and reach into the defined cryogenic regime ($<$123K). Importantly they operate with no mechanical vibrations, magnetic/electric fields or moving mechanical/liquid/gas components and are ideal refrigeration solutions in many difficult/sensitive situations eg. optomechanical, space, sensing experiments \cite{Sheik-Bahae:1320634, Seletskiy:2012bw,Hehlen:2013iy,Nemova:2015kv}. In addition, spin properties of diamond defects have recently attracted much attention owing to their long spin coherence times $T_2$. Since $T_2$ times increase at low temperatures, a new way of cooling diamond is important for applications where long spin relaxation times are needed. Microscopic diamond crystals are highly biocompatible and can be functionalised to attach to specific biologically important ligands \cite{biocomp}. 
Cryosurgery and cryotherapy is a method to destroy diseased or harmful tissues within the body and involves cyclic freezing and thawing of the cells involved \cite{cryo3, cryo1, cryosurgery}. Developing optical methods of cooling nanoscopic diamonds  located in diseased cells to cause cell death may  provide a novel route towards targeted cryosurgery.

In the following we will consider, via a primarily theoretical investigation of the capability of optical refrigeration using diamond, and in particular, using Nitrogen (NV) or Silicon vacancy (SiV) defects in diamond. By estimating via experiment and theory, the emission and absorption cross sections for these two defects over a range of wavelengths we are able to estimate the optical cryocooling power based on a suitable two-level model. We predict that cooling can be achieved using both defects however the NV defect promises very significant cooling due to its high quantum efficiency.  We consider the optical cryocooling of NV/SiV doped microdiamonds (MD), and find substantial temperature changes depending on the irradiation laser power, wavelength and quantum efficiency of the defect. We show that it is possible  to measure the temperature change $\Delta T$, in a MD (diameter $<200\,\mu$m), particularly if it is held in 3D in a trap (e.g. optical or electrodynamic trap). There are various means of inferring the MD temperature either via temperature dependent spectroscopic features e.g. the Zero-Phonon-Line width or the diamond Raman line width. Another signature when the microdiamond is trapped in liquid is the temperature dependence of the particle's diffusion rate.

 \section{Model} Models of optical refrigeration include a four-level \cite{SheikBahae:2009kq}, and two-level model \cite{Luo:1998uw}. We note that  the efficiency of optical cryocooling is degraded when there are any non-radiative routes for the optically excited state  to decay through. The latter will mostly heat the material. In the following we find that indeed the achieved temperature change is very sensitive to the overall quantum efficiency of the emitter. We will focus on the two-level model of \cite{Luo:1998uw}. In this model one considers  a coherent laser beam of circular cross section (radius $r_s$ and area $\alpha_{eff}$), with power $P$, and wavelength $\lambda$, which passes through a cylinder (diameter $D$, length $L$), of  cryocooling material containing an ensemble of two level systems with a defect number density $N=n_\downarrow+n_\uparrow$, consisting of defects in the optical ground(excited) state $n_\downarrow(n_\uparrow$), with optical decay rate $\gamma_{rad}$. Assuming the two-level systems  exhibit wavelength dependent absorption and spontaneous emission cross sections $\sigma_{abs}(\lambda)$ and $\sigma_{se}(\lambda)$, one has a saturation intensity $I_S(\lambda)=hc\gamma_{rad}/(\lambda \sigma_{abs}(\lambda))$, and mean fluorescence emission wavelength $\lambda_F=\int\,\lambda\, \sigma_{se}(\lambda)\,d\lambda/\int\,\sigma_{se}(\lambda)\,d\lambda$. Considering the steady state occupation and assuming the material is optically thin over the range of wavelengths of interest, one obtains the cooling power as \cite{Luo:1998uw},
 \begin{equation}
 P_{cool}=NL\alpha_{eff}I_S \frac{\sigma_{abs}(\lambda/\lambda_{F^*}-1)}{1+\sigma_{se}/\sigma_{abs}+\alpha_{eff}I_S/P}
 \;\;,
 \label{Kool1}
 \end{equation}	
 where $\lambda_{F^*}=[1/\lambda_F-\kappa/(hc\gamma_{rad})]^{-1}$, incorporates any heating $\kappa$, generated by non-radiative decay or other processes in the diamond e.g. Raman scattering. In the case of the NV defect the branching of the decay of the optical excited state through non-radiative paths can be reduced greatly by exciting from the $|m_s=0\rangle$ ground state. Since one can optically pump the NV into this state with high efficiency before initializing the cooling we first assume $\kappa\sim 0$. We also expect that the long wavelength irradiation required for cryocooling will continue to pump the NVs into this ground state. We now consider a roughly spherical diamond particle which is held in a 3D trap. To estimate the change in temperature (the cooling region of the laser and diamond is cylindrical),  one must determine the thermal load and this depends on the physical surroundings. In \cite{Luo:1998uw}, the primary thermal load arises from blackbody absorption of   ambient room temperature radiation by the cold particle. Assuming the change in temperature of the material $\Delta T=T_{mat}-T_{amb}$, is small as compared with the ambient temperature then one has 
 \begin{equation}
 P_{load}\approx -4 A \epsilon_{eff} \sigma_B\, T_{amb}^3\,\Delta T\;\;,
 \label{load1}
 \end{equation}
 where $\sigma_B$ is the Stefan-Boltzmann constant, $\epsilon_{eff}$ is the effective emissivity of the cylinder, i.e. the fraction of energy emitted or absorbed relative to that emitted by a thermal black body, and $A$ is the surface area of the spherical diamond particle. We underestimate $\Delta T$ by assuming $\epsilon_{eff}=1$. Combining (\ref{Kool1}) and (\ref{load1}) we obtain
\begin{equation}
\Delta T=\frac{N\alpha_{eff}I_S}{4\pi D\epsilon_{eff}\sigma_B\,T_{amb}^3}\frac{\sigma_{abs}(1-\lambda/\lambda_F)}{1+\sigma_{se}/\sigma_{abs}+\alpha_{eff}I_S/P}\;\;.
\label{deltaT1}
\end{equation}
In \cite{Luo:1998uw} researchers studied the optical cryocooling of ZBLANP glass doped with  Yb${}^{3+}$ ions in a cylindrical optical fiber with diameter 250-$\mu$m and demonstrated cooling of the fiber to 21 K below room temperature. Crucial towards achieving this are the detailed spontaneous emission and absorption cross sections. In the following section we will examine the case of NV defects in diamond.

\section{Optical Cryocooling of NV defects in diamond}\label{NV1}
\begin{figure}
\centering
\includegraphics[width=.89\linewidth]{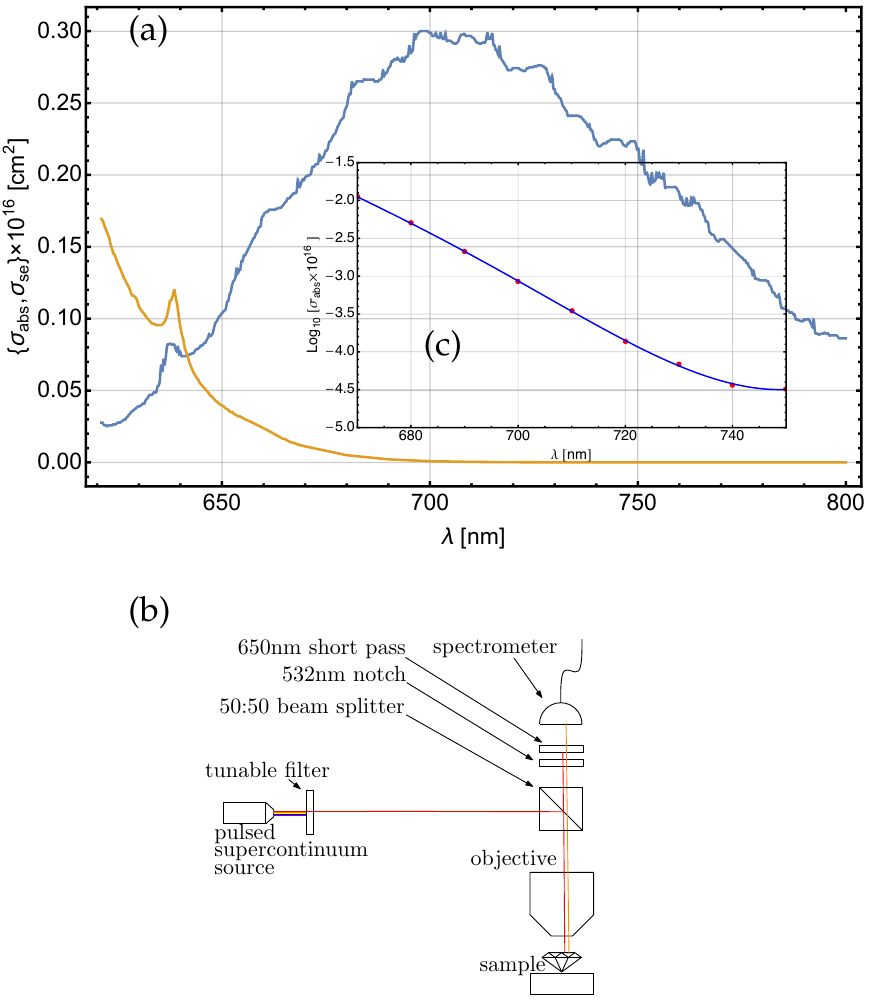}
 \caption{(Color online) (a) Room temperature absorption (yellow), and emission (blue), cross sections for NV diamond over the spectral range of interest for anti-Stokes emission. (b) Schematic of the detailed absorption cross-section measurement setup (see text in Appendix A). (c) [Inset] Measured absorption cross section displayed on a Log plot [red markers, blue is fitted quartic guide to eye].} 
 \label{Fig1}
\end{figure}

We now consider a spherical diamond crystalite of diameter $D$, pumped by a laser beam of cross sectional area $\alpha_{eff}$. As mentioned above the cooling power crucially is dependent on the absorption and emission cross sections. Precise measurements of these cross section are difficult due to the uncertainties in estimating the defect number density $N$. However room temperature NV${}^-$ absorption  and emission cross sections/coefficients as a function of wavelength have been reported in the literature  \cite{Acosta:2011wv,Han:2009iz}. Using the absorption coefficient measured in \cite{Acosta:2011wv}, and their sample's  defect number density $N\sim 2\times 10^{24}\,{\rm m}^{-3}$, we obtain an estimate for  $\sigma_{abs}(\lambda)$. However \cite{Han:2009iz}, reports the emission cross section  $\sigma_{se}(\lambda)$ directly  yielding a mean emission wavelength $\lambda_F\sim 721$ nm. Thus optical refrigeration will require pumping of  NV defects with  $\lambda> 721\,$nm.  The absorption cross section $\sigma_{abs}(\lambda)$ for $\lambda> \lambda^*=670\,$nm becomes quite small and since the cooling efficiency will depend strongly on $\sigma_{abs}$,  we performed experiments to measure this quantity with high precision for $\lambda> \lambda^*\,$ (see Appendix A). These measured cross sections for   $\lambda>\lambda^*$, were combined with those in the reported  literature for $\lambda< \lambda^*$ \cite{Acosta:2011wv,Han:2009iz}, to obtain Fig \ref{Fig1}.

\begin{figure}
 \centering  
\includegraphics[width=\linewidth]{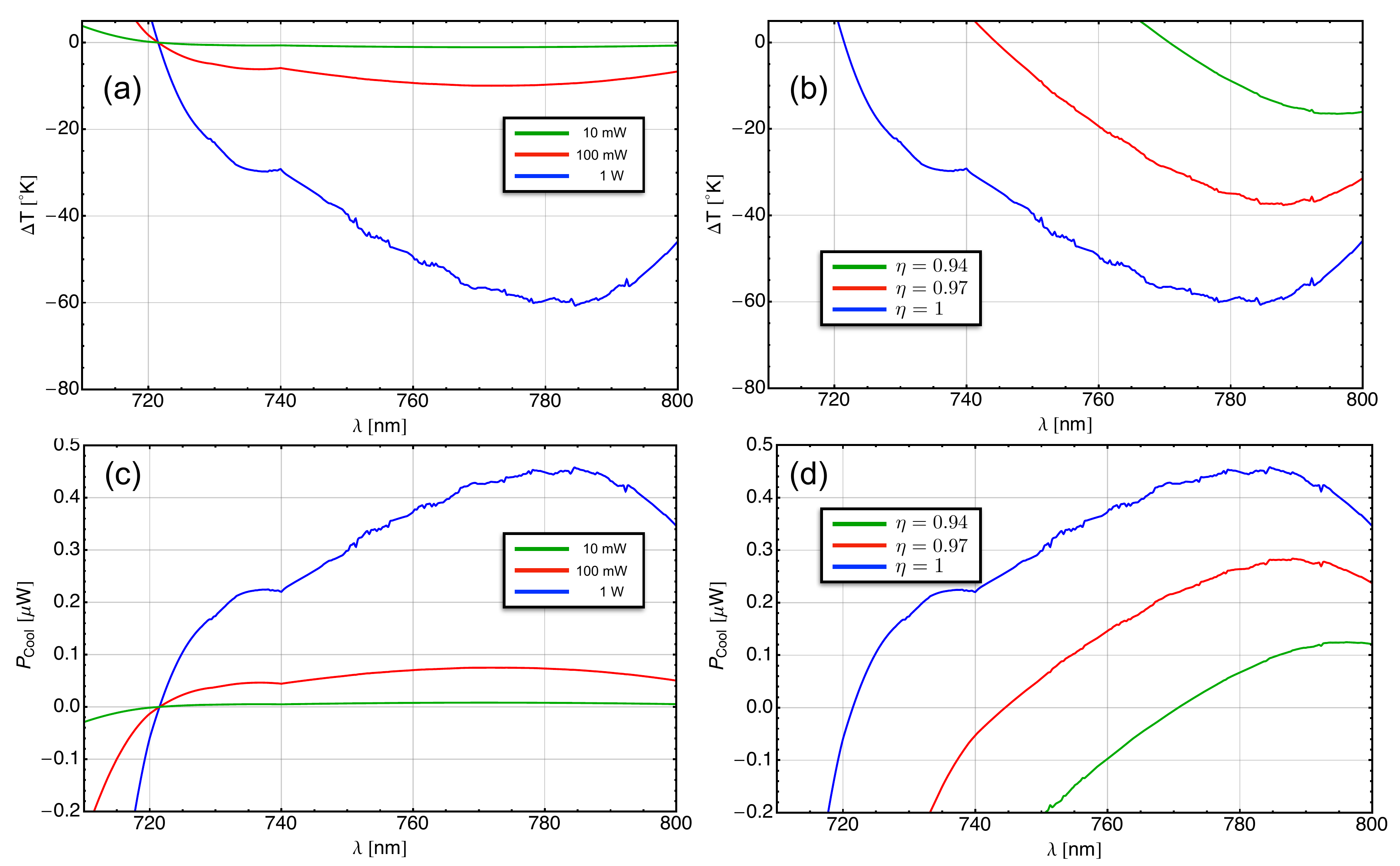}
 \caption{(Color online) Cooling of a NV-doped $20\,\mu$m diameter diamond crystalite suspended in vacuum. (a) We plot the equilibrium decrease in temperature $\Delta T$, of the diamond from ambient as a function of the cooling laser wavelength $\lambda$, for three choices of the laser power $P$ assuming unit quantum efficiency $\eta=1$; (b) $\Delta T$ for a power of $1\,$W, as a function of quantum efficiency $\eta$ assuming all the non-radiative decay processes heat the diamond; (c) Cooling power $P_{cool}$, for same parameters as in (a) assuming unit quantum efficiency; (d) Cooling power as (b) where we vary the quantum efficiency.
} \label{NVCooling}
\end{figure}

We now consider a spherical diamond of diameter $D$, containing an ensemble of NV${}^-$ defects with $N\sim 2.65\times 10^{24}\,{\rm m}^{-3}$ ($\sim$ 15ppm), pumped by a laser beam of spot radius $r_s=5\,\mu$m, wavelength $\lambda$, and power $P$. Taking the optical excited state decay rate $\gamma_{rad}^{-1}=12\,{\rm ns}$, and unit emissivity $\epsilon_{eff}=1$, we now must consider the ambient thermal environment of the diamond. This ambient environment will provide a heating(cooling) rate which will attempt to  bring the diamond back to the ambient temperature. The largest temperature difference will be obtained when the diamond is the most thermally isolated from its surroundings i.e. heats/cools due only to blackbody radiation.  This situation can only be achieved if the particle is levitated in vacuum and the levitation mechanism (e.g. long wavelength optical tweezers \cite{Gieseler:2012bia}), does not heat/cool the particle.  In the case of vacuum trapping  we can use (\ref{deltaT1}), with a $D=20\,\mu$m diamond, to obtain the cooling results shown in Fig \ref{NVCooling}(a). We note that one can obtain substantial cooling with 100 mW laser irradiation but that  levitating nanodiamonds in high vacuum is experimentally  challenging (optical trapped NDs in air at atmospheric pressures have been reported by \cite{Neukirch:2013kg}, while using parametric cooling can trap nanodiamonds in high vacuume \cite{Gieseler:2013aa,lev1}). Researchers initially found that optical levitation of diamonds in air causes extreme heating of the nanodiamonds due, possibly, to burning \cite{Heat1Neukrich, Heat3Hoang}. Recently however experiments in nitrogen atmospheres with high purity diamond show that nanodiamonds can be optically levitated with no heating or burning \cite{MorleyBurn2016}. In the above we assumed that the optical cycling of the NV defect contained no non-radiative transitions which could heat the diamond. In fact the {\em quantum efficiency}, defined as the ratio of radiative to total decay rates $\eta=\gamma_{rad}/(\gamma_{rad}+\gamma_{nrad})$, of NV${}^-$ in nanodiamonds have been shown to vary considerably between different NDs \cite{Frimmer:2013ef}. We can take heating due to non-radiative processes  into account in Eqn (\ref{deltaT1}), through setting $\lambda_F\rightarrow \lambda_{F^*}\equiv [1/\lambda_F-\kappa/(hc\gamma_{rad})]^{-1}$, where $\kappa$ quantifies the non-radiative heating and effectively increases the mean emission wavelength $\lambda_{F^*}\ge \lambda_F$ \cite{Luo:1998uw}. We now consider the worst case scenario where all the non-radiative decay processes heat the diamond and set $\kappa=\gamma_{nrad}\,(hc/\lambda_F)$, to obtain $\lambda_{F^*}=\lambda_F\times \eta/(2\eta-1)$.  In Fig. \ref{NVCooling}(b) we graph the equilibrium temperate change for a range of quantum efficiencies and observe that the cooling is very sensitive to the actual quantum efficiency of the NV defects.  We note that NV defects with quantum efficiencies ranging between 0.7-0.99 have been measured \cite{Gruber:1997tn,Rittweger:2009jx,Waldherr2011,Schietinger:2009jq,ISI:000319856300020}.
Placing the diamond crystalite on a thermally insulating substrate (Pyrex glass), or optically trapping the diamond in liquid e.g. water, will reduce this temperature difference between the diamond and its environment. In Figs. \ref{NVCooling}(c), (d), we plot the corresponding cooling powers obtained for identical situations as in Fig. \ref{NVCooling}(a), (b). The cooling powers obtained are not as large as those potentially achievable using rare earth materials however the latter  are typically highly cytotoxic and cannot be used in biological applications \cite{toxic}.

\section{SiV${}^-$ Defect Cryocooling}\label{SiV1}
\begin{figure}
 \centering  
 \includegraphics[width=.8\linewidth]{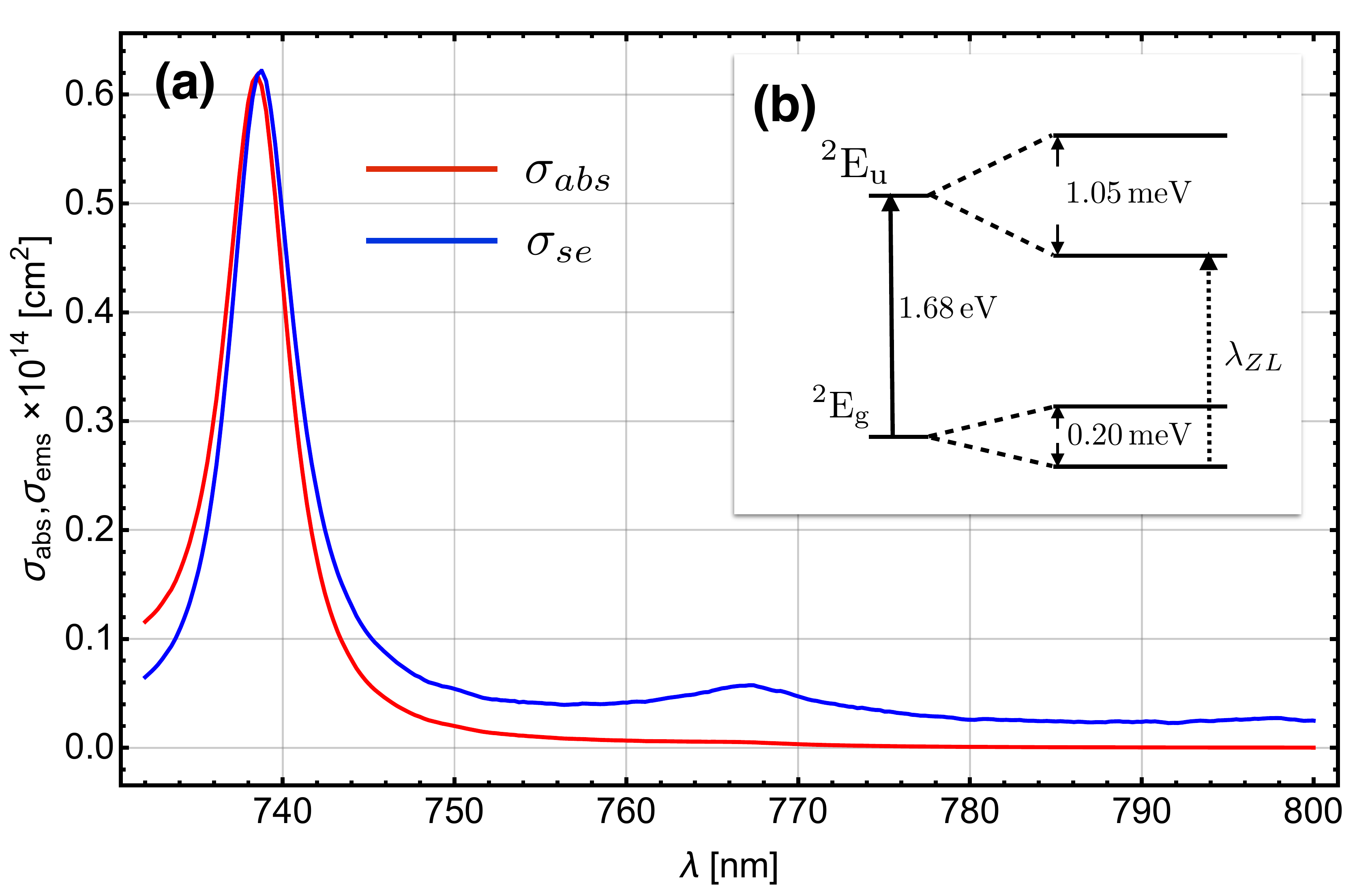}\\
 \caption{(Color online) (a) Room temperature emission and absorption cross section of the SiV${}^-$ defect derived from photoluminescence spectra \cite{Dietrich:2014fg}, and McCumber theory. (b) electronic structure of the SiV defect \cite{Rogers:2014fy}.
} \label{SiVCrossSections}
\end{figure}

The negatively charged Silicon Vacancy defect SiV${}^-$ \cite{Goss:1996ue, Anonymous:DuzrTTX9, Hepp:2014hb, Rogers:2014fy, Sipahigil:2014gh, Rogers:2014if}, has attracted much interest recently for potential use in quantum information. It's photophysical and electronic properties have been studied \cite{Neu:2012up, Rogers:2014fy}, and this defect has a particular advantage for cryocooling due to its narrow emission spectrum and very high emission and absorption cross section with an excited state lifetime of $\tau_{rad}=1.2$ ns at room temperature. The room temperature photoluminescence spectra from an individual SiV${}^-$ defect has been reported in \cite{Dietrich:2014fg}.  The defect possesses a ground and excited state doublet (each of which are doubly degenerate) \cite{Rogers:2014fy}, as shown in Fig \ref{SiVCrossSections}(b).  We now examine its potential applicability in optical cryocooling. Using  Fuchtbauer-Ladenburg theory we can obtain the emission cross section as, 
\begin{equation}
 \sigma_{se}(\lambda)=\frac{\lambda^5\,I(\lambda)}{8\pi n^2c \tau_{rad}\,\lambda_F\,\Sigma}\;\;,
 \end{equation}
 where $\lambda_F\equiv \frac{1}{\Sigma}\int \lambda I(\lambda)\,d\lambda$, and $\Sigma\equiv \int I(\lambda)\,d\lambda$, and the refractive index of diamond $n=2.4$. From this it is possible to estimate the absorption cross section using reciprocal relations first derived by McCumber \cite{McCumber:1964uv} (see (39) in \cite{Mungan:1999vd}  or (9) in \cite{jonchere:2015iu}, and section 3.2.2 in \cite{TAD:2008wea}), as
 \begin{equation}
 \sigma_{se}(\lambda)=\sigma_{abs}(\lambda)\frac{Z_l}{Z_u}\exp\left[\frac{(E_{ZL}-hc/\lambda)}{k_BT}\right]\;\;,
 \end{equation}
 where the partition functions for the $Z_u$, upper and $Z_l$, lower doublets are given as 
 \begin{equation}
 Z_{x}= \sum_{k=1}^2\,d^x_k\,e^{-E^x_{k}/k_bT}\;\;,
 \end{equation}
 where $E^x_{k}$ are the energies of the $k'$th level in the $x=u$ upper, or $x=l$ lower, doublet with degeneracies $d^x_k$. Following \cite{TAD:2008wea} we obtain
 \begin{equation}
 \frac{\sigma_{se}(\lambda)}{\sigma_{abs}(\lambda)}=Z_{ratio}\times\exp\left[\frac{\epsilon-hc/\lambda}{k_B T}\right]\;\;,\label{Vemuru}
 \end{equation}
 where
 \begin{equation}
 Z_{ratio}^{-1}=\frac{(1-e^{-\Delta E_l/k_BT})}{(1-e^{-\Delta E_u/k_BT})}\times \frac{(1-e^{-d_u\Delta E_u/k_B T})}{(1-e^{-d_l\Delta E_l/k_BT)}}\;\;,
 \end{equation}
 where $\Delta E_l=0.2$ meV, $\Delta E_u=1.05$ meV, $d_l=d_u=2$ for the SiV defect and $\epsilon=hc/\lambda_{ZL}$. Using (\ref{Vemuru}) we find the absorption cross section shown in Fig \ref{SiVCrossSections}(a). We note that different from the case of the NV${}^-$ defect, there is significant spectral overlap between the emission and absorption cross sections and therefore we may expect this defect to exhibit a greater cooling capability.

 \begin{figure}
\centering
\includegraphics[width=.75\linewidth]{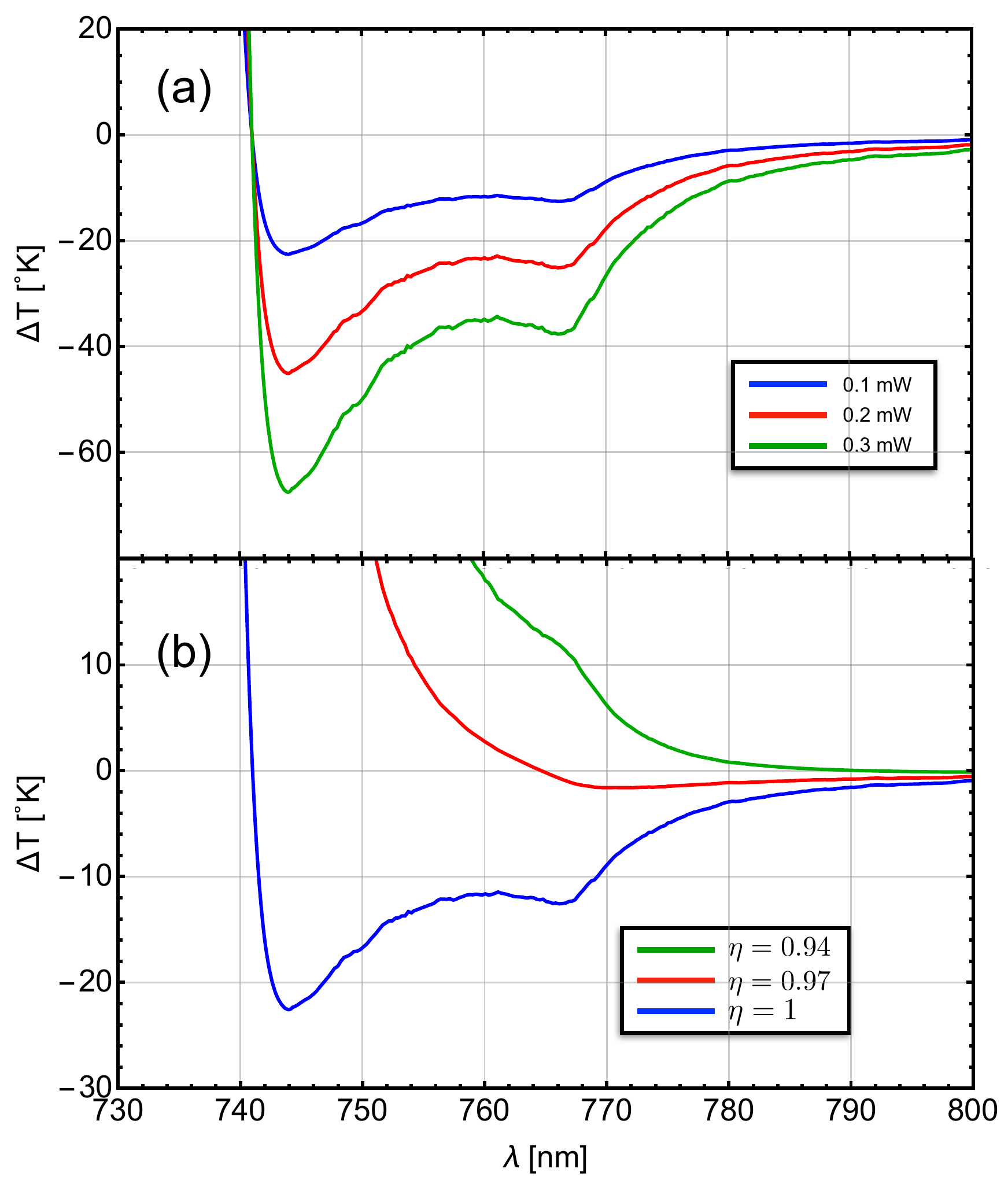}
 \caption{(Color online)  Cooling of a SiV-doped $20\,\mu$m diameter microdiamond suspended in vacuum. (a) We plot the equilibrium decrease in temperature $\Delta T$, of the diamond crystalite from ambient as a function of the cooling laser wavelength $\lambda$, for three choices of the laser power $P$. (b) $\Delta T$ for a power of $0.1\,$mW, as a function of quantum efficiency $\eta$ assuming all the non-radiative decay processes heat the diamond. We note that experimentally one study has reported values for $\eta\sim 0.1$ \cite{Neu:2012up}.
} \label{fig3}
\end{figure}
Using the cross sections shown in Fig \ref{fig3}, and choosing a SiV defect number density of $N\sim 2.65\times 10^{23}\,{\rm m}^{-3}$, ($\sim$ 1.5 ppm), in a relatively large microdiamond of diameter $D=20\,\mu$m pumped by a laser with spot size $r_s=5\,\mu$m,  levitated in vacuum with radiative heating from the ambient room temperature surroundings we obtain the equilibrium temperature change from Eqn (\ref{deltaT1}), as shown in Fig \ref{fig3}(a). Here we estimate $\lambda_F=741$ nm from \cite{Anonymous:DuzrTTX9} (Fig 11). From Fig \ref{fig3}(a) we find that this defect displays tremendous cooling power. The large $\Delta T$ predicted by Eqn (\ref{deltaT1}), assumes that the cross sections do not significantly change with temperature. From \cite{Dietrich:2014fg} it is known that the emission cross section of SiV at 8 K display much narrower features and this temperature narrowing in the cross sections is ignored in Eqn (\ref{deltaT1}). For a fixed pump wavelength we expect that ignoring this sharpening overestimates the actual $\Delta T$ expected. In addition we have also assumed no non-radiative heating of the microdiamond i.e. unit quantum efficiency $\eta=1$. Initial experiments to determine the quantum efficiency of the SiV${}^-$ defect in nanodiamonds reports quite low values $\eta< 0.1$ \cite{Neu:2012up}, at room temperature. Bulk values however have been reported as large as 0.63 \cite{Riedrich14}. In Fig \ref{fig3}(b), we plot $\Delta T$ as a function of $\eta$ assuming $\lambda_{F^*}=\lambda_F\times \eta/(2\eta-1)$, i.e. all  non-radiative decay heats the microdiamond. 
To date there have been only one or two experimental studies on the quantum efficiency of the SiV defect either in nanodiamond or bulk form. Much remains to be understood regarding the factors that influence the observed values. However our study shows that a dramatic reduction in cooling power is associated with even a small decrease in the quantum efficiency from unity. We thus conclude that unless factors influencing the quantum efficiency of SiV defect are better understood, this defect has little possibility to display optical cryocooling.

\section{Measuring the temperature of the diamond crystalite}
To verify the optical cryocooling of the diamond crystalite one must be able to accurately measure the temperature of the micron sized diamond. This is a very challenging task. In the above we have assumed no direct thermal contact between the levitated diamond in vacuum and its surroundings (purely radiative contact), and thus the only methods to measure the temperature of the diamond is to either study the spectrum of the  blackbody radiation emitted by the diamond  or by analyzing temperature dependent features \cite{Kucsko:2013cz,doi:10.1021/nl401216y}, of the emitted photoluminescence of the defects themselves. 
The former method is unrealistic as the blackbody powers involved are extremely small while the latter requires continual spectral analysis.

To focus on the latter, particularly in the case of an NV doped diamond, there is strong dependence on the characteristics of the NV's ZPL and temperature \cite{Thermo1, Thermo3, Thermo2}, and this can be used to determine the particle's temperature with high precision. However we also now explore an alternative method of sensing the change in temperature which may prove particularly useful in cases when the particle is trapped in liquid, as may be the case in bio-compatible applications.

\begin{figure}
\centering
\includegraphics[width=.5\linewidth]{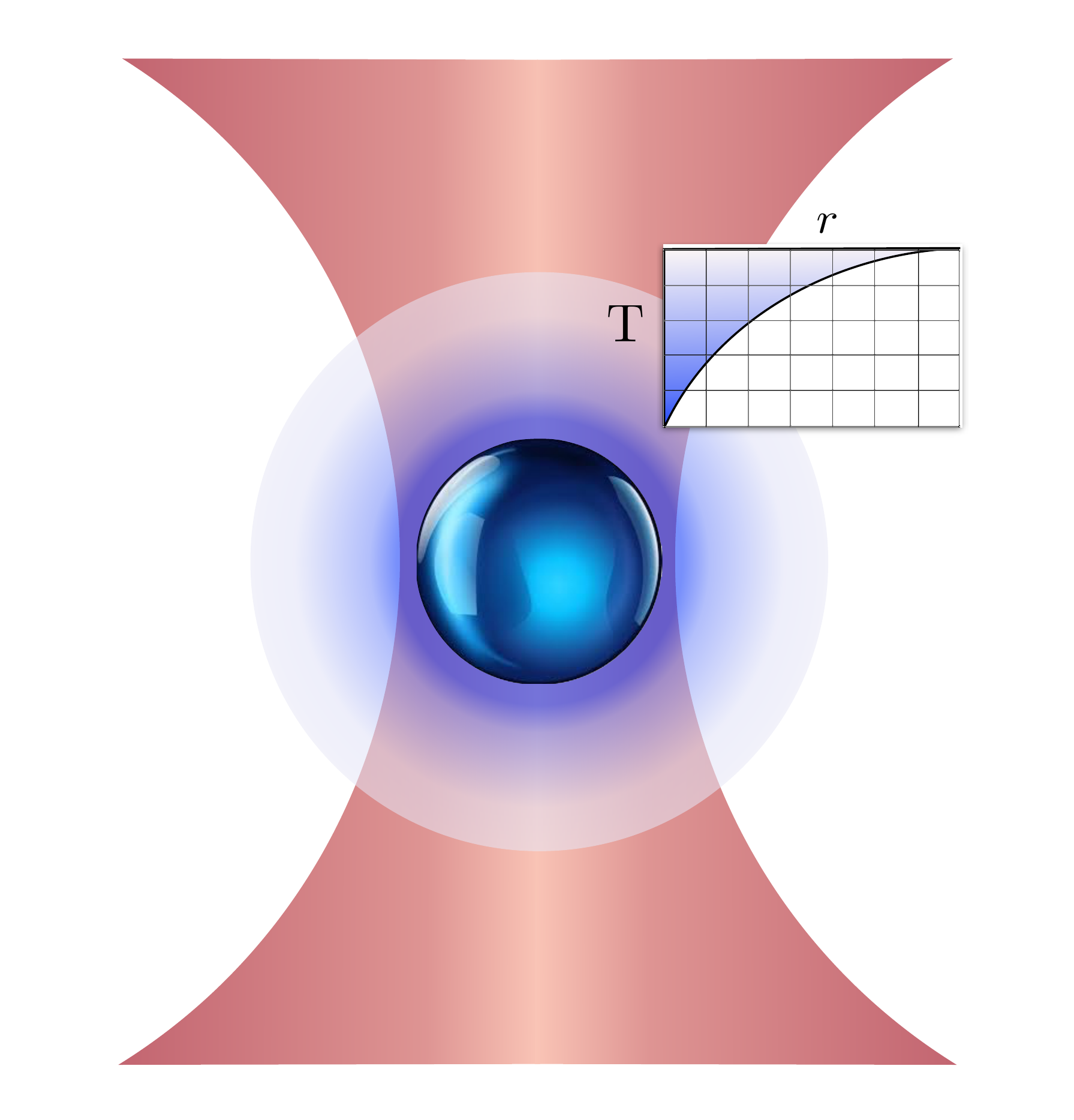}
 \caption{(Color online)  Schematic illustration of estimating the temperature change of a diamond crystalite undergoing optical cryocooling. We perform optical cryocooling on a diamond (blue sphere) held in optical tweezer trap (shaded red) in liquid (white). The radial temperature gradient in the liquid (see schematic inset), formed surrounding the diamond creates a situation of {\em cold Brownian motion}, with an associated diffusion constant $D_{CBM}$. When compared with the Brownian diffusion constant with no cooling $D_{amb}$, we find that the ratio $\chi\equiv D_{CBM}/D_{amb}$ varies significantly in response to optical cryocooling.
} \label{schematic}
\end{figure}

We consider the motional dynamics of the diamond crystalite optically trapped in liquid, the latter held at room temperature but where the diamond is out of thermal equilibrium e.g. at an increased or decreased temperature than the ambient liquid (see Fig \ref{schematic}). This temperature difference causes a change in the motional dynamics of the diamond held in the optical trap via a change in the Brownian Motion stochastic force. By using the theory of so-called {\em Hot/Cold Brownian Motion} \cite{Rings:2010iya, Rings:2011gj, Chakraborty:2011kk, Millen:2014dd, Falasco:2014iq, Roder:2015gy, Anonymous:SXMWFefi}, we predict that the diffusion constant of the diamond will be altered significantly as a function of the diamond's temperature. As the diffusion constant can be accurately measured using position/velocity tracking the diamond's temperature can be easily estimated.

\begin{figure}
 \begin{center}
\setlength{\unitlength}{1cm}
\begin{picture}(7,11)
\put(-1,5.1){ 
\includegraphics[width=.85\linewidth]{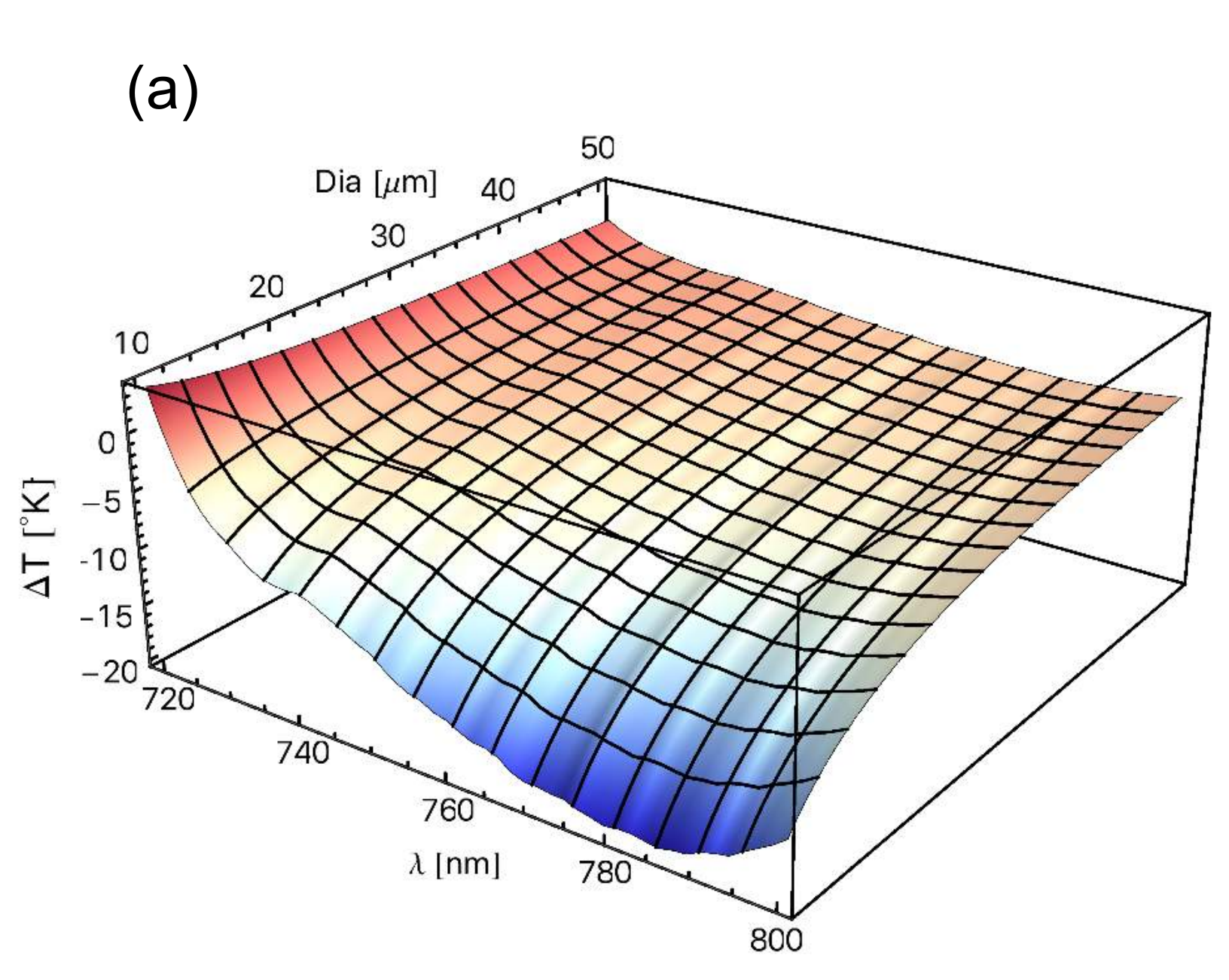}}
\put(-1,-.6){ 
\includegraphics[width=.85\linewidth]{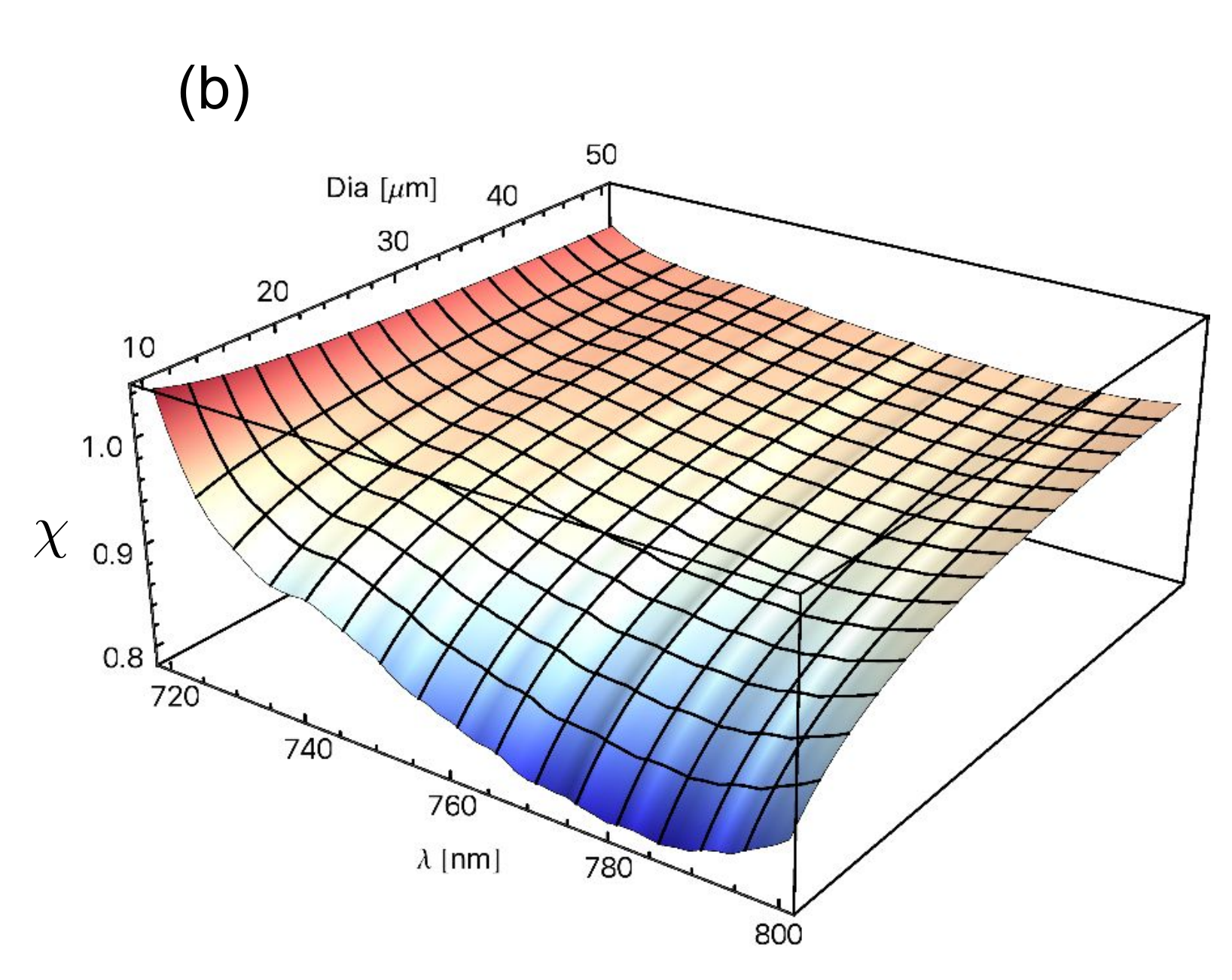}}
\end{picture}
\end{center}
 \caption{(Color online)  Temperature change and modification of Brownian motion diffusion constant in a NV doped diamond crystalite with unit quantum efficiency. (a) NV diamond as in section \ref{NV1}: temperature change as a function of illumination wavelength and diamond diameter for a cooling power of 1 W. (b) dependence of the diffusion constant ratio 
$\chi=D_{CBM}/D_{amb}$. } \label{water}
\end{figure}

We consider the micron sized diamond crystalite to be held in optical tweezers in the liquid solvent $\rm D_2O$ (this solvent is chosen for low absorption at relevant optical wavelengths). In a typical optical tweezers setup the fluctuating position/ velocity of the diamond is recorded in the signal of a quadrant photodiode. If the diamond is not in thermal equilibrium with the surrounding liquid and is held at a higher or lower temperature,  a radial temperature gradient in the liquid  is established and the Brownian motion of the particle is modified. This modification is known in the literature as {\em hot or cold Brownian Motion} \cite{Rings:2010iya, Rings:2011gj, Chakraborty:2011kk, Millen:2014dd, Falasco:2014iq, Roder:2015gy, Anonymous:SXMWFefi}. In particular, in the case of cold Brownian motion (CBM), the diffusion constant is related to the effective CBM temperature of the particle $T_{CBM}$, and CBM Stokes drag $\gamma_{CBM}$, via $D_{CBM}=k_B T_{CBM}/\gamma_{CBM}$ \cite{Anonymous:f8Q2jpqQ}. Here $T_{CBM}=T_{amb}+5\Delta T/12$, where $\Delta T$ is the change in temperature of the particle from the ambient and the Stokes drag $\gamma_{CBM}(T)=6\pi\,R\,\eta_{CBM}(T)$, where $R$ is the radius of the particle. The temperature dependent CBM viscosity $\eta_{CBM}(T)$, is related to the viscosity of the liquid solvent at room temperature $\eta_0$ by
\begin{eqnarray}
\frac{\eta_0}{\eta_{CBM}(T)}&\approx&
1+\frac{193}{486}\left[ \ln\left(\frac{\eta_0}{\eta_\infty}\right)\right]\left[\frac{\Delta T}{(T_{amb}-T_{VF})}\right]\nonumber\\
&-&
\left[ \frac{56}{243}\ln \left(\frac{\eta_0}{\eta_\infty}\right)-\frac{12563}{118098}\ln^2\left(\frac{\eta_0}{\eta_\infty}\right)\right]\times\nonumber\\
&&\left[\frac{\Delta T}{(T_{amb}-T_{VF})}\right]^2\;\;,
\end{eqnarray}
where for the solvent $\rm D_2O$, $\eta_\infty=3.456\times 10^{-5}\,	{\rm Pa\cdot s}$, $T_{VF}=160$\,K and $\eta_0\equiv\eta(T_{amb})$, where the temperature dependence of the solvent viscosity is modeled as $\eta(T)=\eta_\infty\,\exp\left[A/(T-T_{VF})\right]$, where $A=478.7$\,K for $\rm D_2O$ \cite{Anonymous:f8Q2jpqQ}.
To accurately estimate the diffusion constant one must calibrate the quadrant photodiode signal to the actual spatial displacements of the particle but we instead look at the ratio $\chi=D_{CBM}/D_{amb}$, which requires no calibration. We note now that the thermal load is no longer given by (\ref{Kool1}), since there is now convective heating of the the cold particle and  instead of (\ref{deltaT1}), we have
\begin{equation}
\Delta T=\frac{N\alpha_{eff}I_S}{\pi D(4\epsilon_{eff}\sigma_B\,T_{amb}^3+h_{cv})}\frac{\sigma_{abs}(1-\lambda/\lambda_F)}{1+\sigma_{se}/\sigma_{abs}+\alpha_{eff}I_S/P}\;\;.
\label{deltaT2}
\end{equation}
where $h_{cv}$ is the convective heat transfer coefficient for water $h_{cv}\sim 30\, {\rm W/(m^2\,K)}$ \cite{HeatCoeff}. In Fig \ref{water}(a)-(b) we plot the temperature change $\Delta T$ and diffusion ratio $\chi$ for a range of diamond diameters assuming  quantum efficiency $\eta=1$  for  NV doped (as  in section \ref{NV1}). From these graphs we observe that the NV doped diamond exhibits a significant decrease in the ratio $\chi$ particularly for smaller diameter diamonds and this should be readily observable in an experiment.

\section{Conclusion}
The development of all optical cryocooling has potentially numerous applications ranging from situations which require low-power and low-vibration, through to refrigeration in difficult environments eg. space, high-pressure, electrical systems where a liquid coolant is problematic. Traditional materials for optical cryocooling include rare-earth crystals and semiconductors however such materials are typically highly toxic in biological environments. In contrast, in this work we have found that diamond doped with Nitrogen Vacancy defects show potential for optical cryocooling. 
We proposed the optical cryocooling of levitated diamond crystalites which, due to their thermal isolation, may exhibit large temperature reductions using optical cryocooling. 
For our work we found that cryocooling requires the defects to possess high quantum efficiencies and NV defects with such high levels of quantum efficiencies have been experimentally demonstrated in nanodiamonds \cite{Femius13}, and post-selecting nanodiamonds to achieve a uniformly high quantum efficiencies is feasible.   

Another route to increase the optical quantum efficiency which is particularly suited to the NV defect is to enhance the ZPL emission using the Purcell effect via coupling the defects to an optical cavity.  Optical cryo-cooling is based on the principle that more phonons are absorbed than created. Since the NV centre has a strong phonon-sideband and hence produces lots of phonons during its radiative decay, cooling effects only start at wavelengths considerably above the ZPL and are highly dependent on quantum efficiency. This could be fundamentally improved by a change of the relative emission pathways associated with the Purcell-enhancement of  a particular emission wavelength of the NV centres inside a cavity. Purcell enhancement is, by now, an established technique 
\cite{Su:08,Su09,  Kaupp13, Moller15, Wolf15}. An enhancement of the ZPL has a twofold advantage: The emission on the phonon sideband is reduced relative to the ZPL which does not create any phonons and the optical life-time of the NV is shortened, strengthening the radiative decay relative to any non-radiative pathways, which do not change in their decay rate, i.e. increasing the quantum efficiency. While NV diamond inside a cavity sets some limitations in terms of applicability it might provide strong advantages for cooling: lowering the effective excitation wavelength above which a cooling effect starts, which in turn increases the absorption cross-section, i.e. the cooling effect at the same power as well as improving the cooling effect through an effective increase of the quantum efficiency. This might be an interesting avenue for further improvements.

Thus  we have found that the optical cryocooling of NV doped diamond crystalites is viable and there may be potential use for this in biocompatible applications e.g. in cryosurgery or cryotherapy. The use of other defects in diamond with higher Debye-Waller factors and near unity quantum efficiency would prove even better candidates for optical cryocooling of diamond.
\section*{Appendix A}
The absorption cross section of the negatively charged NV centre is well known only for the excitation wavelength $\lambda=532$nm. For cryocooling it is important to measure the absorption cross section as a function of wavelength for $\lambda > \lambda^* = 670$ nm without relying on typically inaccurate estimates for the number of emitting NV centers and the laser spot size. We achieved this by comparing the emission from the NV$^-$ centers (zero phonon-line emission at 637nm) for different excitation wavelengths and relying on the literature value for 532nm excitation \cite{Chapman:2011be}. We ensured that the excitation power was consistently at 200$\mu$W for all high wavelengths and accounted for the reduced power of our 532nm reference measurement. We also ensured that the excitation power is low enough to not reach any saturation within the measurement to guarantee linear behavior with excitation power and comparability across all excitation wavelengths. Different excitation wavelengths were created by a supercontinuum source with a tunable filter, see Fig. \ref{Fig1} and focused onto a single crystal diamond sample with very high NV centre density to ensure a good signal. A 650nm short pass and 532nm notch blocked the sensor from reflections of the excitation light, which were much stronger than the signal. The incoming light was then directed to a spectrometer and measurements with an acquisition time of 1s were taken. We compared the signal at the NV$^-$ zero-phonon line and the relative signals as a function of wavelength were then linearly scaled to an absorption cross-section by equating the reference intensity, measured with 532nm excitation, to the literature value of 0.95$\times 10^{-16}$ cm$^2$ \cite{Chapman:2011be}. Note that this measurement is more precise than an absorption measurement because out of all the possible absorption mechanisms we select only that of the NV centers by measuring the emission of NV centers rather than the reduction in the transmitted light. 

\section*{Acknowledgements}
The authors thank B. Gibson (RMIT), and K. Xia (MQ), for useful discussions and L. McGuinness (U Ulm), for discussions and provision of a NV-dense diamond sample. 
This work was supported by the Australian Research Council Centre of Excellence in Engineered Quantum Systems EQUS (Project CE110001013) and ARC  Discovery Project DP130104381.

%

\end{document}